# Redirecting Flows – Navigating the Future of the Amazon

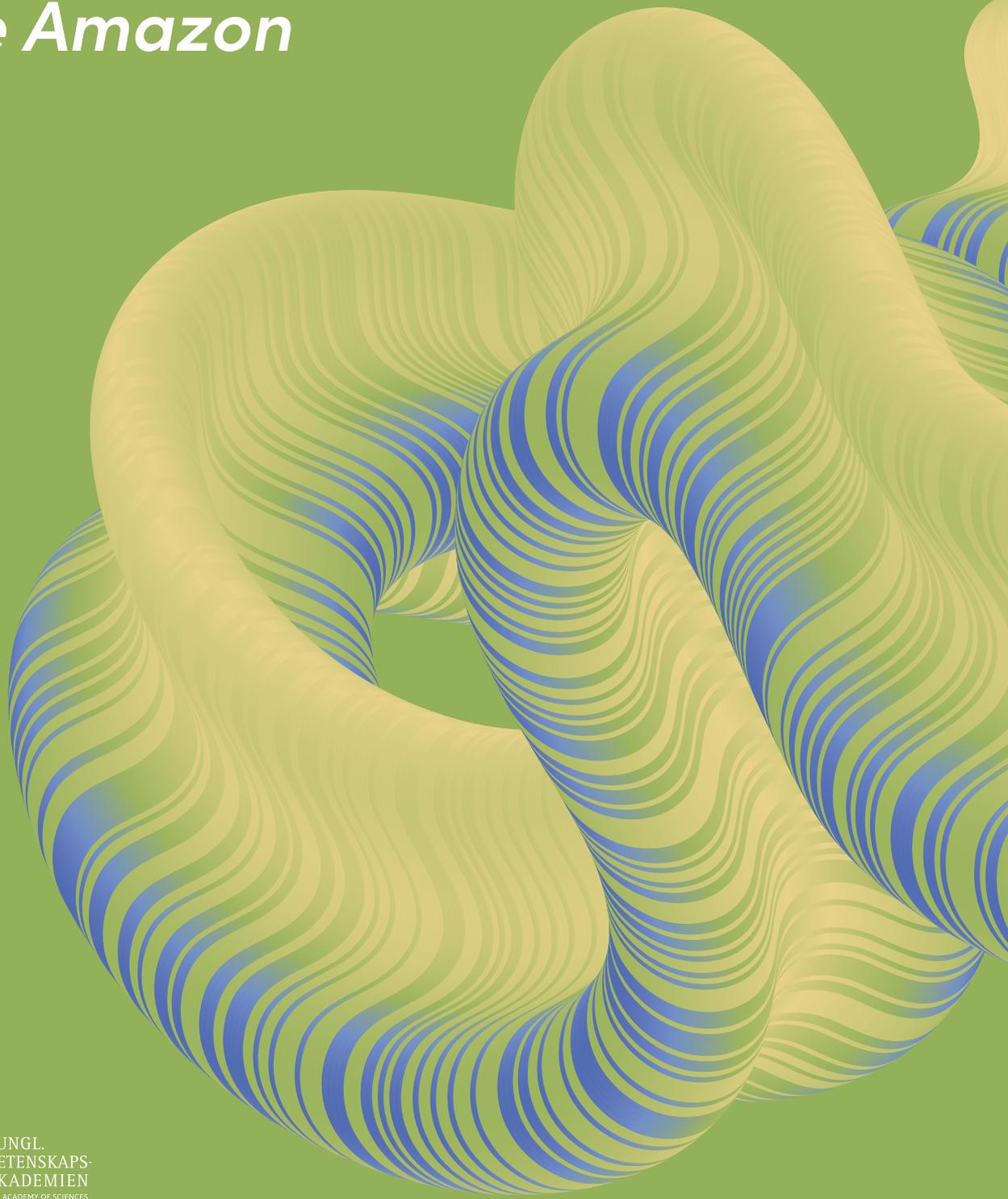

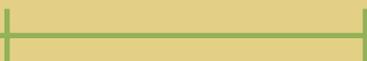

**Contributors**

This report would not have been possible without the contributions, critical reflections, and comments from the authors listed below. Any mistakes are the sole responsibility of the main authors, and the views presented in the report do not necessarily reflect those of each contributor, their respective organizations, nor funders.

Chapter 1. Victor Galaz, Megan Meacham

Chapter 2. Garry Peterson, Megan Meacham

Chapter 3. Victor Galaz

Chapter 4. Juan C. Rocha

Chapter 5. David Armstrong McKay

Chapter 6. Paula Sanchez, Bianca Voicu

Chapter 7. Svante P. Persson, Regina Cervera, Constanza Gomez Mont

Chapter 8. Adrian L. Vogl, Mary Ruckelshaus

Chapter 9. Ana Paula Aguiar, Taís Sonetti-González, Minella Martins, Francisco Gilney Bezerra, Aldrin Perez-Marin

Chapter 10. Megan Meacham, Victor Galaz

This report has been edited by Victor Galaz and Megan Meacham. We thank Mary Ruckelshaus and Svante P. Persson for their critical reflections and comments, and Bianca Voicu for the coordination of the report.

**Funders and Contributing organizations**

The report is supported by Marianne & Marcus Wallenberg Foundation (grant number MMW2017.0137) and "Networks of Financial Rupture – how cascading changes in the climate and ecosystems could impact on the financial sector" funded by Formas' National climate research programme. The work presented here also builds on research undertaken within the Stockholm Resilience Centre, the Natural Capital Project at Stanford University, the Inter-American Development Bank (IDB), MISTRA Finance to Revive Biodiversity (grant number DIA 2020/10), The Beijer Institute of Ecological Economics at the Royal Swedish Academy of Sciences, and the Global Economic Dynamics and the Biosphere program (GEDB).

*This report should be cited Galaz, V. and Meacham, M. (eds., 2024). Redirecting Flows – Navigating the Future of the Amazon. Report. Stockholm Resilience Centre, Stockholm University. https://doi.org/10.48550/arXiv.2403.18521*

Design and illustrations by Elsa Wikander and Jerker Lokrantz at Azote.

# *Executive Summary*

*The Amazon Basin, and the Latin America and Caribbean (LAC) region, stands at a critical juncture, grappling with pressing environmental challenges while holding immense potential for transformative change through innovative solutions. This report illuminates the diverse landscape of social-ecological issues, technological advancements, community-led initiatives, and strategic actions that could help foster biosphere-based sustainability and resilience across the region.*

- Biodiversity serves as the foundation for economic prosperity and human well-being, providing essential ecosystem services such as pollination, carbon sequestration and water purification. Loss of biodiversity due to human activities, such as deforestation and wildlife trading threatens food security, water quality, health, and security, impacting millions worldwide.

- The Amazon rainforest, home to around half of the world's remaining rainforests and a significant portion of global biodiversity, faces substantial threats from deforestation and climate change. Evidence of a potential "Amazon dieback", where the rainforest could transition to a drier, more degraded state due to reduced moisture recycling and increased droughts, is a major concern. Dieback could be triggered by factors such as deforestation reaching a threshold of around 20-25% of the original rainforest territory or global warming reaching 3-4°C.

- Interactions among tipping points and Earth Systems cascading effects can propagate irreversible changes across ecosystems, affecting climate feedback loops and amplifying environmental risks beyond the boundaries of the Amazon rainforest. If wide-scale dieback were to occur, the Amazon rainforest could release significant amounts of $CO_2$ into the atmosphere, exacerbating global warming and regional climate impacts, such as changes to the hydrological cycle and precipitation patterns across South America.

- The financial sector plays a significant role in driving biodiversity loss by funding extractive activities that result in direct and systemic socio-economic risks, which can trigger instability across businesses, markets and livelihoods. Moreover, the financial sector is not shielded from the impacts of biodiversity loss. A potential 'tipping' of the Amazon rainforest into a savannah can be considered a "green swan event", posing significant financial risks with wide-ranging social, economic, and ecological implications.

- Extreme weather events, such as droughts triggered by climate change, are already affecting the Amazon Basin and hydropower infrastructure, leading to the risk of stranded assets and disruptions in electricity generation. Nearly 1,000 existing or planned hydroelectric power plants in South America are at risk of further decrease in precipitation, with financial implications for investors, including publicly-listed companies and large asset managers.

- Action to limit warming to 1.5°C or well below 2°C, as outlined in the Paris Agreement, is crucial, but not sufficient, to minimizing the risk of Amazon dieback. Restoring degraded forest, shifting to sustainable land-use practices, and empowering Indigenous peoples and local communities are key strategies for improving Amazonian resilience to climate change and halting deforestation. This requires collaboration across biodiversity science, accounting, regulation, and business practices to promote sustainable financial practices and nurture biodiversity for long-term prosperity, with a few strategies already being rolled out across the region.

- Equity, loans and bonds provide pathways for investors to influence corporate policies and actions. U.S.-based asset managers play a significant role in investments associated with key drivers of deforestation, especially in Latin America. Mobilizing financial influence can complement governmental efforts by engaging with the corporate sector to advance financial transparency and support a resilience and a planetary health agenda.

- Novel technologies aimed at mitigating environmental issues and enhancing nature-based solutions, could revolutionize approaches to preserve biodiversity and combating climate change. By filling critical data gaps, so-called "nature tech" can empower decision-makers with the information needed for effective land use planning, conservation, and restoration efforts, thus attracting investment in sustainable initiatives. However, several challenges must be overcome to fully leverage this potential, including limited funding, unequal access to technology, fragmented collaboration, and insufficient capacity building. Ethical considerations are paramount to ensure responsible technology usage, equitable benefits distribution, and mitigation of unintended consequences.

- While technology and infrastructure can partially replace nature's benefits, they are often costlier and fail to provide multiple co-benefits. Natural Capital Assessments and Accounting (NCAA) have emerged as effective tools for integrating biodiversity and ecosystem services into policy and finance decisions. Through participatory science-policy processes, stakeholders are co-creating actionable, science-based solutions to address pressing challenges. In the Amazon Basin, natural capital approaches are already influencing policy and investment decisions, through building community resilience in the Amazon Headwaters, harmonizing livelihoods in Bolivia's Llanos de Moxos region, and helping prioritize investments in Colombia.

- Lessons from neighbouring regions can also inform action within the Amazon rainforest. The XPaths project engaged stakeholders in a participatory process to identify key challenges and formulate strategic actions for a sustainable future in the semiarid region of São Francisco River basin. Four primary challenges and interconnected strategic actions were identified: environmental education and social mobilization, agrarian reform and territorial demarcation, political empowerment and policy continuity, diversification of the economy. By addressing these challenges holistically, the semiarid region can move towards a sustainable and just future, safeguarding water sources and promoting socio-economic development for generations to come.

# *Table of contents*

# Chapter 1.
## Introduction

*Victor Galaz and Megan Meacham*

Our living planet and the climate system are changing at an unprecedented speed. The way our economies are organized, and capital is allocated, needs to adapt to this new planetary reality to secure a safe and just future for all. These changes need to happen with a profound sense of urgency and based on the best available evidence.

Our ambition is to offer the latest research insights from research exploring the connections between investments, ecosystems and the biosphere (chapter 2 and 3). We focus in particularly on issues related to the Amazon, including the dynamics of abrupt changes in Earth systems (also known as "tipping points") and implications for the region (chapter 4 and 5); the way some of these changes cascade through economic sectors and finance thus resulting in novel financial risks (chapter 6); and on tangible solution pathways to help mitigate and act on such risks (chapter 7-9).

The work presented here is an extension of previous work presented at Stockholm +50, June 2022. We hope that this review offers some new insights and contributes to forceful political action and financial innovation that is truly able to secure a prosperous future for all.

This review is the result of a collaboration between the Natural Capital Project (Stanford University), Stockholm Resilience Centre (at Stockholm University, through the research projects "Cascading financial risks" funded by Formas, FinBio funded by Mistra, and MMW 2017.0137), and the Inter-American Development Bank.

# Chapter 2.
## The connection between biodiversity and systemic financial risks

*Garry Peterson and Megan Meacham*

Economic activity is unweaving the web of life. Finance is a major force directing this economic activity. This chapter explains how biodiversity sustains economics and finance, and how losses of biodiversity threaten the viability and stability of economic activities. The chapter also explores current activities from biodiversity science, economics and regulation that are being advanced to address these risks.

## The Essential Role of Biodiversity in Finance

Finance, economies, and human societies are all embedded within the biosphere (Folke et al., 2016; Dasgupta, 2019). The living world is the ultimate source of human well-being. However, sustainability researchers recognize that the increase in scale, connectivity and speed of the global economy is reshaping the biosphere, and its capacity to reliably support humanity (IPBES, 2019; Nystrom et al., 2018). The planetary boundaries framework focuses on defining safe limits to these modifications (Rockström et al., 2009) and recently there have been attempts to define just limits (Rockström et al., 2023).

International science policy assessments have documented how human existence and quality of life depend on nature in increasing detail. The Intergovernmental Science-Policy Platform on Biodiversity and Ecosystem Services (IPBES) described how nature plays a critical role in providing food, energy, medicines and a variety of materials fundamental for people's physical well-being, social support, and cultural identity. Ecosystems contribute to fundamental functions like pollination, carbon sequestration, water purification, and climate regulation. These ecosystem services have monetary value and are vital for various industries, including agriculture, pharmaceuticals, and tourism (TEEB, 2010). Genetic diversity, an essential component of biodiversity, is crucial for agriculture and medicine, playing a pivotal role in the creation of resilient and high-yielding crops, as well as in drug development (Heywood, 1995). Marine and terrestrial ecosystems are the key sinks for human caused carbon emissions, and currently store about 60% of human caused emissions (Rockström et al., 2021).

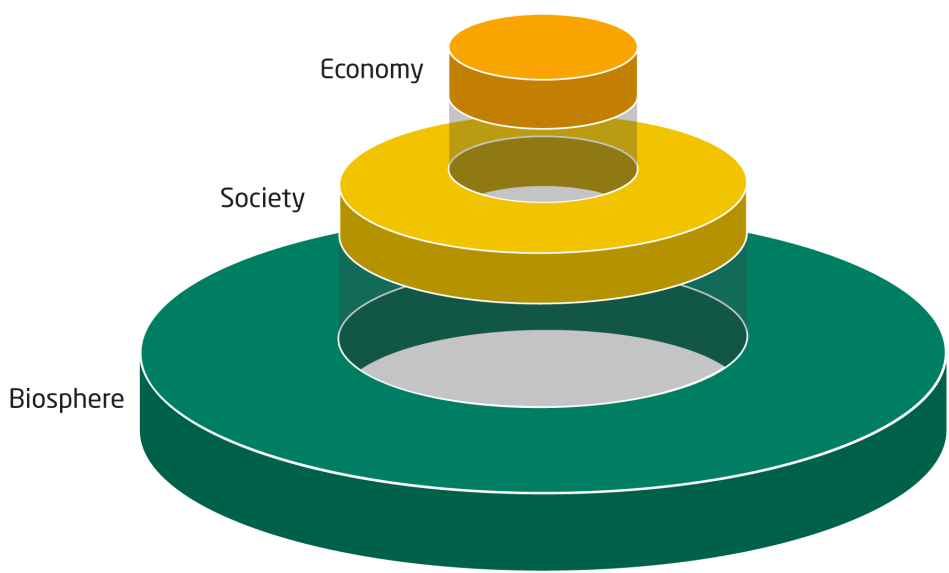

**Figure 1** | The economy and society as embedded within the biosphere, as intertwined parts of the planet. The biosphere serves as the foundation upon which prosperity and development ultimately rest. From Folke et al. (2016).

Biodiversity loss is changing nature by reducing the supply of ecosystem services, reducing the resilience of ecosystems, and reducing the ability of nature to adapt to change. Biodiversity loss undermines key aspects of human wellbeing: food security, water quality, health, and security. By reducing the productivity of agricultural systems, biodiversity loss reduces access to food for millions of people who depend on crops and livestock for their livelihoods. By impairing the ability of ecosystems to provide clean water, biodiversity loss reduces access to safe drinking water for billions of people who rely on natural sources. By creating new opportunities for zoonotic spread of disease, biodiversity loss makes people more vulnerable to infection by exposing them to novel pathogens and vectors. By destabilizing ecosystems and making them more variable and more likely to experience regime shifts, biodiversity loss reduces the predictability and stability of the services that ecosystems provide. Furthermore, by making ecosystems less resilient to extreme weather events, such as floods, droughts, and wildfires, biodiversity loss increases the risk of disasters that can destroy homes, infrastructure, and livelihoods.

The benefits that nature provides people are fundamentally irreplaceable (Dasgupta, 2019). Technology and infrastructure can enhance and replace some of the benefits that people receive from nature, but most of them are more expensive than maintaining nature, incur high future costs, and fail to provide multiple co-benefits. For example, storm surge and coastal protection can be provided by seawalls and dikes, rather than mangroves. However, technology and infrastructure are also costly, often difficult to maintain, and do not provide other ecological benefits such as spawning habitat, carbon sinks. Nature's ecological and evolutionary processes maintain these capacities in the living world and provide nature and people with the capacity to adapt to future changes in the living and non-living world.

## Finance's Impact on Biodiversity and Systemic Financial Risks

The financial sector steers economic activity by providing investment, loans, and insurance. The financial system could be funding the restoration and revival of biodiversity, but it currently does not. Instead, it promotes the expansion of activities that are simplifying ecosystems and driving the loss of biodiversity (IPBES, 2019). Not only are many of these activities bad investments, but these activities can also produce new types of risks and shocks (Dasgupta, 2019). Understanding both the direct and systemic economics risks arising from biodiversity loss becomes crucial for economic resilience and sustainable development.

Nature loss does not only increase the likelihood of extreme events which impact society, it also reduces the capacity of nature and society to cope with and respond to these shocks. While it is difficult to estimate the exact size of economic losses which will result from an eroded natural environment, it is not difficult to estimate some possibilities. For example, nature loss is increasing the risk of zoonotic diseases. The emergence of a disease with greater impacts than the COVID pandemic could disrupt global supply chains, as well as reduce consumer and business activity. The economic toll of the COVID-19 pandemic for the U.S. economy is estimated to reach US$14 trillion by the end of 2023 (Walmsley et al. 2023). These shocks could lead to banking, debt, and currency crises, and the interconnectivity of global financial networks could amplify the effects, transmitting risk from vulnerable areas to other areas. For example, US$4.3 billion was spent preventing and treating malaria in 2016 (Haakenstad et al. 2019). More locally, conversion of natural ecosystems, and in particular the loss of wetlands in urban watersheds could lead to severe urban flooding, leading to substantial property and infrastructure damage, as well as business interruptions. If such losses overwhelm local insurance mechanisms, the financial consequences of such an event could trigger sectoral, local, and national crises, particularly in the banking and debt sectors, highlighting the vulnerability of urban infrastructure to ecological changes and the systemic risks posed to financial institutions that finance and insure such assets.

Many types of economic development are based on subsidies which promote activities that decrease the public benefits from nature (Barbier et al., 2022). For example, the World Bank recently estimated that continued land clearing in Brazil could cost the country US$317 billion annually. While public economic benefits from ecological conservation are estimated to be seven times as valuable as the economic benefits from agriculture, logging, and mining (Hanusch, 2023), the beneficiaries from these activities are different.

Furthermore, the destabilization of ecosystems can produce novel types of shocks or risks for economic actors and the financial system. For example, agriculture is a major sector in Brazil's economy and a key driver of Amazonian deforestation. Amazonian deforestation appears to be reducing rainfall and water availability in ways that are threatening agriculture, water for cities and hydroelectric power generation (Keys et al., 2019a; Leite-Filho, 2021), thus resulting in new material financial risks (see Chapter 6). By destabilizing the biosphere, economic activity can produce new types of so called "Anthropocene risks" that in turn destabilize economic activity but are difficult to forecast and quantify (Keys et al., 2019b; Rising et al., 2022). This vulnerability can trigger cascading effects through the economy, leading to widespread instability, affecting businesses, markets, and livelihoods (Henderson, 2021).

Realization of the adverse impacts of biodiversity loss has led to the emergence of sustainable investment movements. There is increasing understanding of the need for both greening finance, to reduce harm, and financing green, to strategically invest in solutions that enhance biodiversity (World Bank, 2020). Reducing harm involves minimizing detrimental impacts on biodiversity, acknowledging its foundational role in ecosystem services. Financing green entails investing in innovative approaches that align with nature-positive production of goods and services, ultimately contributing to halting and reversing nature loss (Locke et al., 2020).

## Going Forward

The recent Kunming-Montreal Global Biodiversity Framework (GBF) was established as a key step towards stopping biodiversity loss. This framework encompasses 23 global targets to be completed by 2030, with the ultimate goal of living in harmony with nature by 2050. Among these targets, the "30X30" agreement, aiming to protect 30% of land and seas, is a significant step in addressing the main driver of biodiversity loss—land-use change (CBD, 2020). The framework emphasizes the need to shift incentives in global finance and business towards nature-positive actions. It also highlights the importance of considering the perspectives and interests of Indigenous peoples and local communities, promoting a pluralistic approach to biodiversity conservation (IPBES, 2019). Achieving the targets outlined in the GBF is a monumental challenge to the status-quo, and while political awareness and international cooperation around nature are historically high, the resources being committed to this goal are still far too low (Deutz et al., 2020).

The Dasgupta Review (Dasgupta, 2019) suggests that businesses and financial institutions should be required to disclose their dependence and impact on nature. The aim of this increased transparency allows investors and shareholders to assess the nature-related risks in their portfolios, as well as consumers and regulators to assess the activities of companies and investors. There is substantial progress towards this goal. The GBF includes target 15, which requires governments to ensure that large and transnational companies disclose "their risks, dependencies and impacts on biodiversity". There is a wider variety of initiatives in Europe, the US, and Japan to track corporate impact, along with the recommendations from the Taskforce on Nature-Related Financial Disclosures that can be expected to continue to develop and be implemented in some form.

Achieving all of these goals will require substantial change in how the world operates and this will require not only biodiversity science, but new accounting systems, regulations and business practices. Creating effective standards will require collaboration between all these fields to implement, monitor, and evaluate, as well as development of new operating practices and standards over time. Striking a balance that promotes sustainable financial practices and nurtures biodiversity is a shared responsibility, requiring collective efforts from stakeholders across sectors.

## Chapter 3.

# How finance influences the emergence of novel disease risks

*Victor Galaz*

Zoonotic diseases can have devastating effects on human health and societies. The COVID-19 pandemic is a stark reminder of how such emerging and re-emerging infectious diseases (EIDs) at times propagate through transnational trade and travel routes and overwhelm fragile health systems. Unfortunately, such risks to both human and animal healthcare are likely to increase substantially in the near future due to the combined effects of climate and land-use change (Carlson et al., 2022).

There is an increased recognition that human activities including deforestation, the expansion of agricultural land, and increased hunting and trading of wildlife, can be linked to the emergence and re-emergence of such diseases, in particular of zoonotic and vector-borne diseases (Allen et al., 2017; Marco et al., 2020). The production of certain types of high deforestation-risk commodities such as palm oil, have been linked to increased zoonotic spill-over through land fragmentation and habitat loss (Morand et al., 2021).

In today's globalized economy, many large companies rely on obtaining external capital from financial institutions to expand their operations and increase production with notable environmental, ecological and social impacts in both land- and seascapes. This means that financial investments and institutions such as banks, pension funds and multilateral development banks may indirectly affect ecological dynamics by funding – and benefitting economically from – extractive activities that create new patterns of interactions between pathogens, non-human animals, and humans (Galaz et al., 2023).

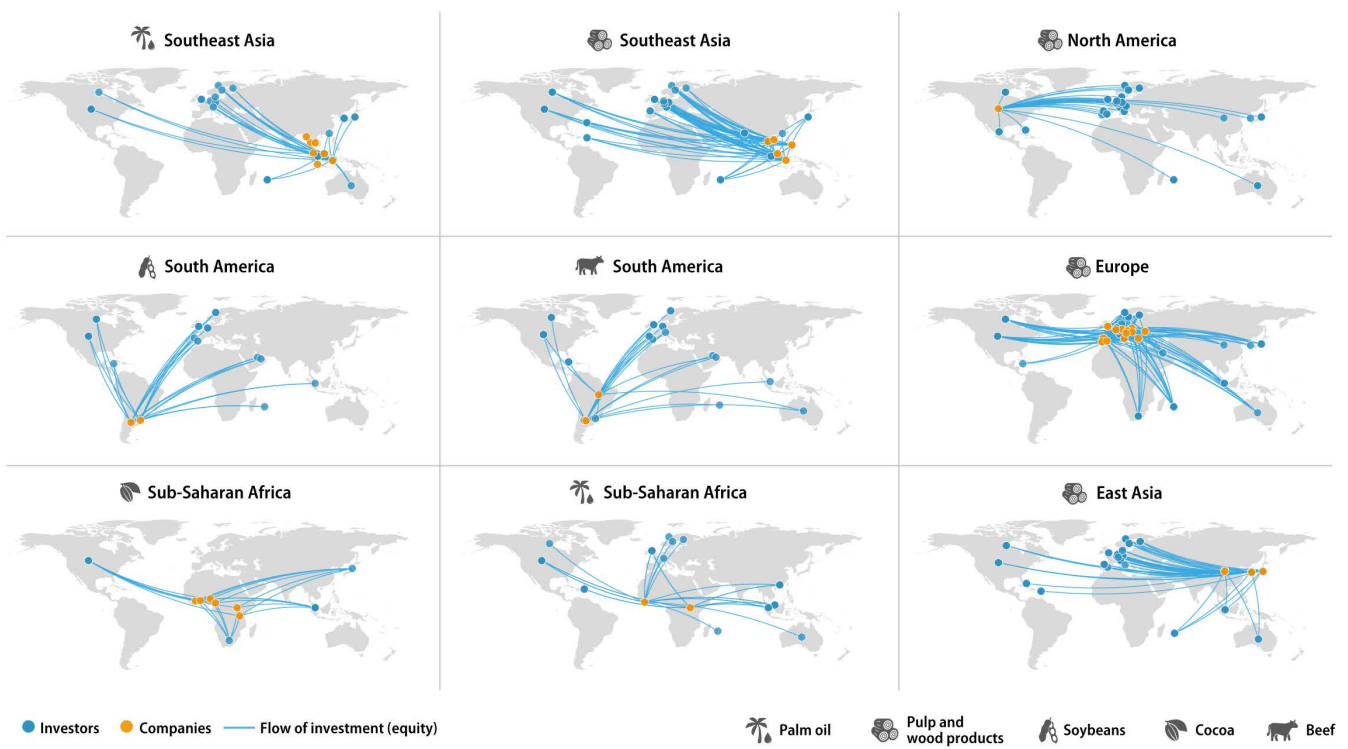

**Figure 1 | Global connections of investments through equity.** Financial investments shape the biosphere, and as a result also emerging and re-emerging disease risks through investments in economic sectors associated with anthropogenic land-use changes in known zoonotic disease "hotspots". The figure includes N=54 companies and shows the global characteristics of such investments in the nine selected regional case studies, as well as the respective investment size through equity in USD. Purple nodes are where companies and investors overlap geographically. Note that the figure is a simplified data-based animation based on (Galaz et al., 2023).

For example, consider financial institutions which hold significant shares in companies that cultivate maize, rice, sugar cane and soybeans for international markets. Through their voting rights as owners, such institutions indirectly hold influence over corporate activities that drive the conversion of landscapes that help maintain rodent populations and rodents' breeding sites, which in turn increases the rodent zoonotic disease spread through inhalation of infected aerosols particles stemming from the rodents' urine, faeces, or saliva (e.g., hantaviruses, see Galaz et al., 2023).

## Financial influence – but where?

Financial instruments such as equity, loans and bonds provide pathways for investors interested in influencing the policies and actions of companies. Financial institutions such as pension funds for example, engage in direct dialogues with corporate management, use their voting influence at corporate annual general meetings, and sometimes threaten to divest as a way to influence companies to act on issues that are central to their interests. When successful, such engagements can have considerable "downstream" effects when large companies choose to use their dominating market position and globally spanning supply chains to advance sustainability and climate ambitions (Folke et al., 2019).

But where in the world is such financial influence at all possible? A recently published study (Galaz et al., 2023) analyzes the possibilities of such influence by tracing equity investments in companies and regions in the world where EIDs risks have been shown to be particularly high and associated with deforestation risk prone commodities. Figure 1 shows the overall investments patterns in six different regions of the world, and two cases in Latin America associated with deforestation-risk commodities.

There is a consistent large role (and potential influence) for U.S.-based asset managers through their diverse ownership in key sectors associated with anthropogenic land use change (Galaz et al., 2023). However, there are important regional differences, especially for Latin America (Figure 2). International attempts to leverage financial influence for planetary health thus not only has to consider dominating global investors, but also has to be adapted to specific regional ownership patterns. Hence, while the prevention of EIDs risks requires global cooperation, progress in mitigating global risks can also be made through strategic alliances between a smaller subset of countries (e.g., Aakre et al., 2019), and through other centrally placed companies and private sector actors (Folke et al., 2019).

The unequal nature of global commodity trade and financial investments often leads to financial institutions investing in companies operating in countries with higher corruption, inequality, and/or weaker rule of law. This observation reemphasizes the need to develop investor policies and engagements that consider limited government capacities.

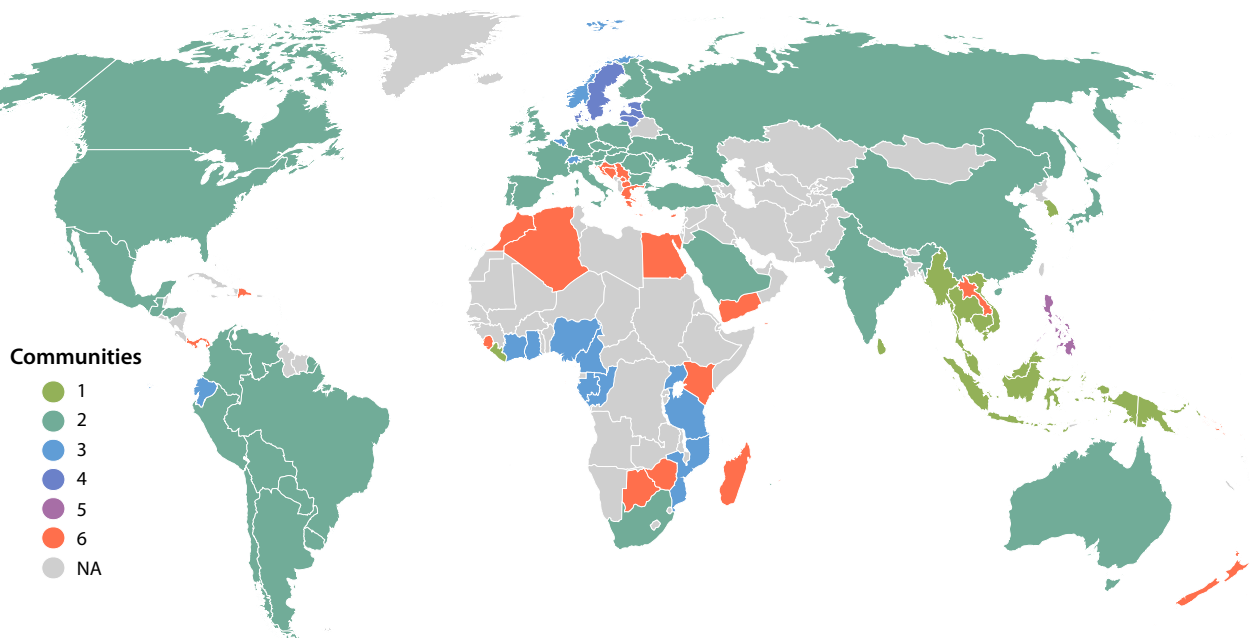

**Figure 2** | The figure shows the different constellation of countries who collectively could help mitigate EIDs risks linked to deforestation risks. Each community shares company headquarters and financial entities associated to economic activities in regions of the world with elevated EIDs risks. From Galaz et al., 2023, Figure 4A.

On January 20th, 2022, the International Monetary Fund raised its forecast for the economic costs of the COVID-19 pandemic on the global economy to US$12.5 trillion through 2024 (Reuters, 2022). These numbers show the large economic impacts and material financial risks created by emerging and re-emerging diseases, and the tangible economic incentives investors and governments have to address such diseases proactively. It can also be seen as one illustration of the possible cascading effects that result from financial investments as they impact on ecosystems, only later to feedback on the economy and on the financial system itself (Sanchez et al., 2022).

Financial influence can – and should - be mobilized to complement the work assumed by governments and international organizations by building alliances with investors with similar interests and using that collective influence to engage with the corporate sector in ways that advance financial transparency of EID-risks, and support the development and implementation of corporate policies aligned with a planetary health agenda.

# Chapter 4.
# *Why Cascading Shifts matter for investors*

*Juan C. Rocha*

The latest report from the International Panel for Climate Change brought the attention of scholars and policy makers to the impending risk of tipping points (IPCC, 2021). Simply put, tipping points are points on a critical parameter, such as temperature, that, once reached, cause a system to flip from one configuration to another. These points are really hard to measure and predict. Nevertheless, we know they exist, and the scientific community has for decades been assessing their risks and uncertainties.

Writing in Science, David Armstrong McCay and colleagues, reported an update on climate tipping points and their risk in 2022 (Armstrong-McKay et al., 2022). They find, for example, that the Amazon rainforest can tip at a 3-4°C increase in average temperatures (see Chapter 5). According to climate models such levels of temperature are reachable in the 2100 century. But recent observations have shown that trees are already reaching such a threshold (at the forest canopy, see Doughty et al., 2023). The Amazon tipping means the forest will no longer be able to play an important role in regulating the climate by capturing and maintaining rich storage of carbon. Instead, it can become a less productive forest or a savanna with detrimental consequences to the species living there today.

It is a concerning fact because the Amazon plays important ecological roles in regulating climate by capturing carbon, as well as acting as a water pump for regions adjacent to the Amazon (Smith et al., 2023; see Chapters 5 and 6). The rainforest creates precipitation that in turn nurtures agricultural landscapes of Latin America and provides water to major urban centers such as Sao Paulo. On the ground, measurements today already show that regions of the Amazon are shifting from being a carbon sink to a carbon source (Gatti et al., 2021).

Thus far scientists have been studying tipping points as independent phenomena. But interactions among tipping points (say, between temperature and precipitation) can exacerbate impacts on human societies. There are two broad classes of interactions. First, interactions of different driving forces on the same system. In the case of the Amazon, consider the many different ways human activities can tip it: by increasing temperature, through human-made fires, or through deforestation. What would happen if we "push" the Amazon on the three drivers at the same time? Scholars believe that the tipping point for deforestation is ~40% with a lower bound of 25% (Nobre et al., 2016). Today we are at 20%, meaning that we are likely to reach deforestation tipping points well before the temperature one.

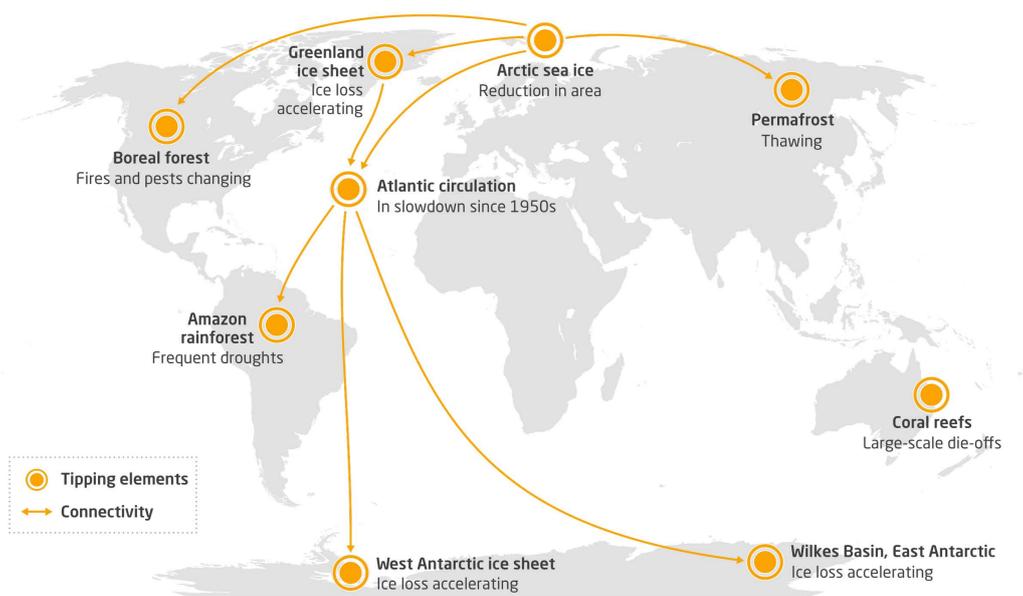

**Figure 1 | Potential global tipping points.**

The diversity of drivers presents challenges and opportunities for managing systems with tipping points (Rocha et al., 2015a). On the one hand, understanding common sets of drivers opens the opportunity to manage multiple ecosystems under similar threats. For example, in managing agricultural activities by reducing the use of fertilizers, fishing, and sewage can prevent multiple regime shifts in coastal systems (Rocha et al., 2015b). In fact, recent studies have shown that many of these drivers occur at local to regional scales of management, so people on the ground can do a great deal already to avoid ecological collapses. Climate change, although important, is not the only threat. By reducing pollution, and addressing urbanization or food production related drivers (e.g. fishing, agriculture, use of fertilizers), many tipping points can be avoided, or at least delayed.

The second type of interaction is known as cascading effects (Rocha et al., 2018). Imagine a system, like a forest, tips in one part of the world. How does it affect the risk that other ecosystems tip? If the Amazon tips, are other ecosystems at risk as, for example, commodity producers look for other areas to expand agricultural production? Or the other way around, what ecosystems, if tipped, would increase the risk of the Amazon tipping as well?

The scientific community in general agrees that such cascading effects are plausible. However, most of the studies have been modelling work with a focus on the climate system, thus only focusing on tipping elements that are affected by increases in temperature. Tim Lenton and collaborators suggest that climate tipping elements are interconnected (Lenton et al., 2021), and subsequent modelling work has clarified the mechanisms through which systems such as the Greenland Ice Sheet, the Atlantic Meridional Ocean Circulation (AMOC), the Arctic sea ice or West Antarctica, can be connected (Wonderling et al., 2021).

However, this body of work ignores the living fabric of the planet: the biosphere. Biodiversity in these studies is often seen as a storage of carbon, but less is understood about its role in regulating nutrient cycles, the water cycle, or in producing the benefits people get from nature. Local to regional shifts in ecosystems might not be strong enough to cause detectable effects on the climate system. For example, shifts in coral reefs affect the habitat for a diverse set of species, undermine recreation opportunities, affect fisheries, and increase food insecurity or lead to coastal erosion. But can changes in coral communities impact the climate?

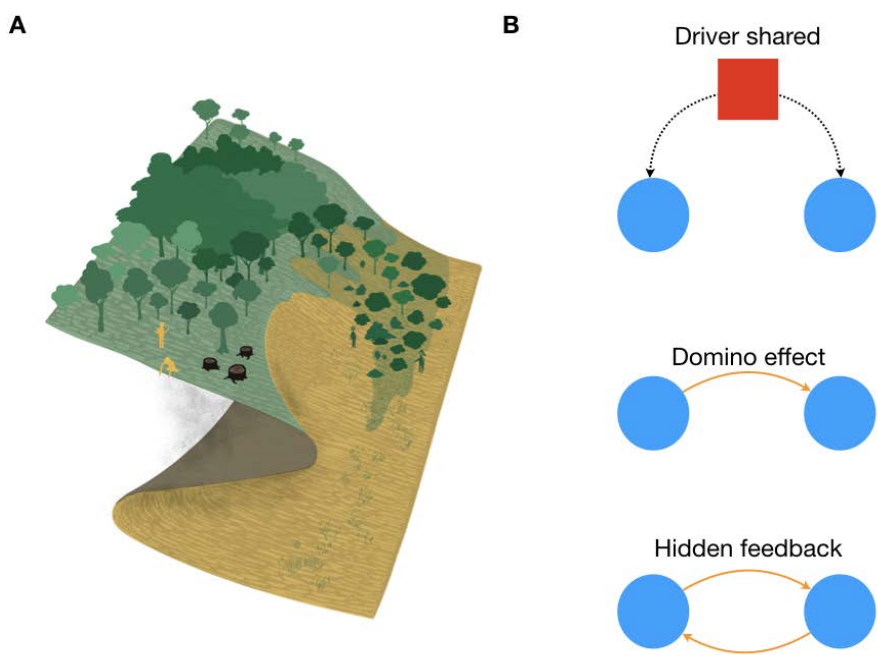

**Figure 2 | Potential cascading effects of regime shifts.** (A) An example of a regime shift from forest to savanna. Regime shifts are typically modeled as the interaction of fast and slow processes that create discontinuous transitions. (B) Driver sharing has the potential to sync two different regime shifts (blue) but not necessarily make them dependent, while domino effects or hidden feedbacks would couple their dynamics by creating structural dependencies (orange). From (Rocha et al., 2018).

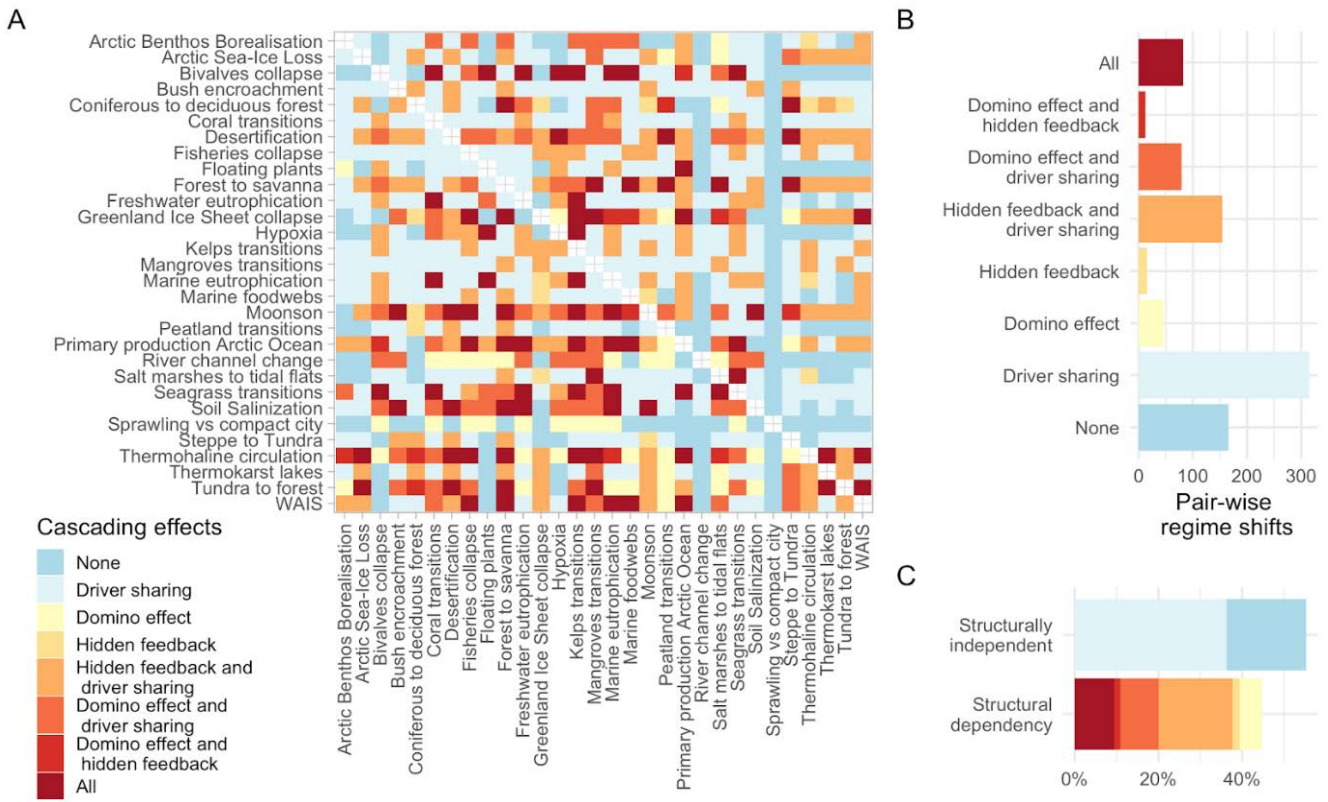

**Figure 3 | Potential structural dependencies between regime shifts.** (A) The three response variables combined show eight different possibilities in which regime shifts can interact through cascading effects. (B) Driver sharing is the most common type found. (C) Domino effects and hidden feedbacks alone or in combination account for ~45% of all regime shift couplings analyzed, implying structural dependence. From (Rocha et al., 2018).

Whether these ecological shifts could amplify or exacerbate climate change is still an open question. Shifts in the frequency of forest fires and forest composition in the boreal forest, could amplify climate change at first, but help capture more carbon in the long term (Mack et al., 2021). Shifts in peatlands and thermokarst lakes can release methane, a strong greenhouse gas that amplifies global warming (Koffi et. al 2020). To what extent the biosphere can amplify or dampen climate change is still a key frontier of research.

Our study analyzed 30 different types of abrupt shifts in ecosystems, and found that about 45% of all possible couplings resulted in cascading effects (Rocha et al., 2018). We showed the different mechanisms that exist and could couple ecosystems. While climate change is one, it is not the only one. Yet the empirical evidence to help assess the strength and likelihood of these couplings, is still missing. Focusing only on climate could be misleading, strangely erring on being too optimistic.

Back to our example of the Amazon rainforest, the temperature threshold is expected in climate models to be reached by 2100s, but the deforestation threshold could be reached in a matter of decades. Fires and current droughts can move the time to tipping even earlier. Similarly, coral reefs can tip already at 2°C, but their probability of recovery is strongly influenced by interaction with other drivers such as fishing pressure, water turbidity, ocean acidification, or the presence of herbivorous fish. Some of these drivers can be managed locally to build resilience against climate change.

A pre-emptive approach that manages the diversity of drivers is imperative. For some of the drivers, local and regional authorities can intervene (e.g., to halt deforestation). We cannot risk crossing arms and wait until the large nations of the world commit to significant actions in reducing emissions and dealing with climate. Opportunities for meaningful actions are available here and now.

## Chapter 5.

# How deforestation and climate change could push the Amazon to a tipping point

*David Armstrong McKay*

The Amazon rainforest is one of the most biodiverse places on Earth, fitting around half of the planet's remaining rainforest and 10% of known species (WWF, 2010) in only 1% of the Earth's surface. Around 17% of the forest has been lost since 1970 (SPA, 2021) and another 17% degraded, mostly from clearance for cattle ranching, soy plantations, logging, and mining. This damage is threatening many rare species and is pressuring the Indigenous peoples and local communities who depend on the forest (Conceição et al., 2021).

As well as direct loss, researchers are also worried that beyond a certain level of deforestation or warming the rainforest could start to retreat on its own, even if deforestation or warming stopped – a tipping point (Armstrong McKay, 2019) known as '*Amazon dieback*'.

This could happen because the Amazon rainforest is partly self-sustaining (Looms, 2017). A rainforest can only grow above a minimum rainfall level, but the Amazon also makes around a third (Staal et al. 2018) to a half (SPA, 2021) of its own rainfall by recycling moisture (Staal et al. 2020) from the Atlantic. Winds transport this moisture further inland in a great 'atmospheric river' where it can be recycled again and again, expanding the area wet enough for rainforest to grow. Moisture recycling also acts like a giant 'air-conditioner', allowing the forest to cool itself (SPA, 2021).

If enough forest is lost due to drought or deforestation in key rain-producing regions though, less rain is recycled, and areas downwind get drier. The Amazon is also seeing more frequent droughts (SPA, 2021), with water levels on the Rio Negro recently hitting record lows (Reuters, 2023) due to this year's El Niño on top of long-term climate change-induced drying. These droughts make wildfires – to which rainforest is particularly vulnerable – more likely. If this pushes these downwind areas below the rainfall level needed for rainforest to survive it could lead to further forest loss, even less moisture recycling, and dieback could cascade through vulnerable parts of the Amazon.

## When could dieback happen?

An estimated 40% of the Amazon (Staal et al., 2020) – mostly in the drier south and east – can tip from a wet rainforest state to a drier, more open degraded forest or savanna-like state (Hirota et al., 2011). Each state sustains itself through feedbacks (such as moisture recycling for rainforest or wildfires for open state), and can tip from one state to the other when pushed beyond climate or forest loss thresholds.

In 2016 Brazilian climate scientist Carlos Nobre led a paper that estimated dieback could be triggered around either 40% deforestation or 3-4°C of global warming (Nobre et al., 2016; Nobre and Flatow, 2019). This matches other studies that found the risk of dieback grows above 2°C (Boulton et al., 2013) and becomes likely beyond around 3.5°C (Armstrong McKay et al., 2022), and would take several decades to a century to fully play out. However, in 2018 Nobre and ecologist Thomas Lovejoy suggested that deforestation-induced tipping could start at as low as 20-25% due to interactions with warming-induced droughts not fully captured by their models, which is worryingly close to the current 17% deforestation level (Lovejoy and Nobre, 2018). Localized dieback may have even started in a few places (Quintanilla et al. 2022), but has not yet spread across larger regions.

There is some uncertainty around model projections and driver interactions though, so these lower deforestation thresholds are based on expert judgement. The IPCC's latest report gave a low likelihood for Amazon dieback this century as it features in only some leading climate models, but project biome large shifts by 4°C of warming (Canadell et al. 2021). Some recent models suggest the forest may be more resilient to climate change (Cox, 2020) than first feared due to high adaptability (allowing it to survive past natural dry periods), and the *Scientific Panel for the Amazon* found a basin-wide threshold was too uncertain to identify (SPA, 2021). Key feedbacks like nutrient limitations or wildfires are not yet well represented in many models though, so these model projections may be over-stable thus underestimating the risks of a transgression of a threshold.

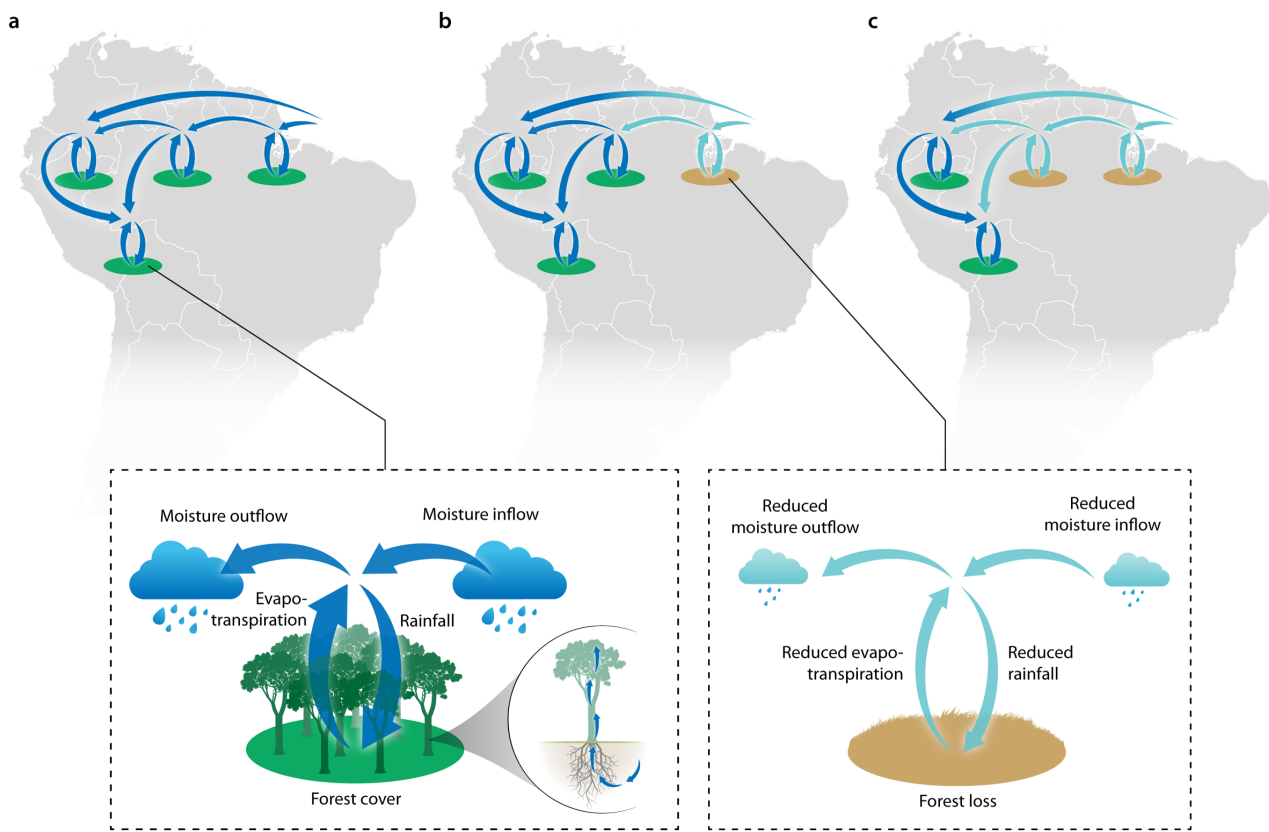

**Figure 1 | The Amazon Rainforest's Tipping Dynamics.** a) Wind from the Atlantic carries moisture westwards over the Amazon, where it falls as rain and is mostly returned to the atmosphere through trees pumping water up and evaporating from leaves ('evapotranspiration'), making the moisture available to move further inland and fall as rain again. b) If drought (or deforestation) drives a shift to a state with fewer trees, this reduces the amount of moisture recycled and therefore the amount of moisture and rainfall downwind too. c) If areas downwind are pushed passed the minimum viable rainfall level for rainforest they can tip into a degraded state too, creating a "dieback" cascade spreading through vulnerable areas. Figure based on Zemp et al. (2017).

Given this, 20-25% deforestation acts as a provisional precautionary threshold that is wise to stay within even if the actual tipping point threshold turns out to be higher, and the Science Panel for the Amazon has called for an immediate moratorium on deforestation in tipping-prone regions (Rodrigues, 2021).

## What's at stake

The Amazon rainforest stores around 150 to 200 billion tons (SPA, 2021) of carbon in its plants and soils, and over the last few decades has absorbed 5-10% of yearly human $CO_2$ emissions from the atmosphere (Cox, 2020). However, the capacity of this 'carbon sink' peaked in the 1990s and is now falling, and combined with degradation the Amazon likely now emits more $CO_2$ than it absorbs (Gatti et al. 2022). This shift is not a tipping point in itself, but means the Amazon has started to amplify rather than counter global warming.

If wide-scale dieback were to start, then it would lock in far more $CO_2$ release over the coming decades. Dieback in the dry south and east could release around 30 billion tons of carbon (McKay, 2022), which is the equivalent of around 3 years of current human emissions, and along with biogeophysical feedbacks could add 0.1°C to global warming over the next century. Even more carbon would be at risk if higher warming makes more forest vulnerable to dieback (Staal et al., 2020). Regional impacts would include extra local warming of up to 1°C as the forest's self-cooling ability is reduced, and reduced rainfall across the Amazon and the Southern Cone (See Chapter 6). Thankfully, the planet's oxygen supply is not at risk though, as it was built up over millions of years (Zimmer, 2019).

Action now is critical for protecting what remains of the Amazon rainforest. Keeping to the Paris Agreement goal of limiting warming to 1.5°C (or well below 2°C) would help to minimize the chance of Amazon dieback, requiring the rapid phase out of fossil fuels and the transformation of global food systems. Restoring degraded or lost forest and shifting to agroforestry can also help restore moisture recycling and recapture some lost carbon, while legally protecting intact forest and empowering Indigenous Peoples and local communities deters deforestation (Boadle and Shumaker, 2019). Together these would help improve Amazonian resilience to climate change, but ultimately the Amazon is not safe until both greenhouse gas emissions and deforestation stop.

# Chapter 6.
# What if the Amazon tips? Exploring the implications for investors

*Paula Sánchez García and Bianca Voicu*

The tipping of the Amazon rainforest into a savannah is considered to be a "green swan event", a potentially extremely financially disruptive event, with wide-ranging social, economic and ecological implications (Bolton et al., 2020). Indeed, impacts on carbon emissions and, hence, global climate stability, or on regional agriculture (Lenton et al., 2019), are often referred to as the main material risks for financial institutions as a result of crossing a tipping point in the Amazon rainforest (Santander, 2021; Stand.Earth, 2022; Svartzman et al., 2021). However, our understanding of how rainforest dieback could lead to large changes in the hydrological cycle within South America, and cascade in ways that amplify financial risks, is limited (Lamb et al., 2022). This chapter brings together insights from climate and hydrological models, with financial data to elaborate the anatomy of such risks.

## Aerial rivers and changes to precipitation patterns in South America

Large-scale rainforest dieback will significantly impact the capacity of the Amazon rainforest to act as a unique hydrological recycling hub. As a result, such a dieback increases the risk of a collapse of the network of water vapor transport which underpins the "aerial rivers" that reach within and beyond the region (Zanin and Satyamurty, 2020). Land use change is expected to significantly impact precipitation patterns in the Peruvian Amazon and western Bolivia, a region which receives 70% of its precipitation originating from the Amazon rainforest (Weng et al., 2018). Furthermore, shifts in the process of water recycling and transport of water vapor through changing patterns of evapotranspiration is predicted to also impact the Atlantic Forest (Ferrante et al., 2023), and the La Plata Basin (Zanin and Satyamurty, 2020). The extent to which different regions in South America might be impacted by a fall in precipitation as a result of land use change in the Amazon rainforest is illustrated in Figure 1.

## Implications for investors

Whether due to a regime shift (see Chapter 4) from tropical rainforest to a savannah, or due to wider system impacts on climate and the hydrological cycle, the alarming levels of deforestation in the Amazon represent material risks to investors and financial stability. The increase in extreme temperatures due to climate change and exacerbated by El Niño have already triggered draughts of unprecedented severity, with direct impacts on the Amazon Basin. In addition to the severe stress on the riverine biodiversity, the record low water levels in the Negro and Madeira Rivers in Brazil saw communities stranded and a large hydropower closed (Knutson, 2023; The Guardian, 2023). As scientists warn of an increase in extreme weather events, this has particularly significant consequences for not only the people of the Amazon, but also in the longer term for the just transition to a low-carbon economy in Latin America as countries in the region are relying on an increasing capacity of the hydropower sector for its green energy sector (McCauley et al., 2023; IEA, 2021).

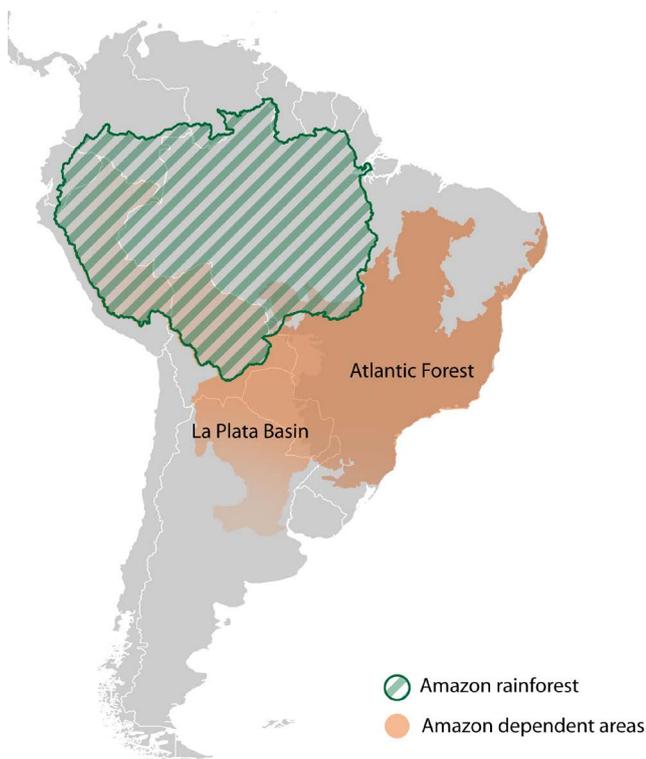

**Figure 1 | Impact of deforestation in the Amazon rainforest on precipitation in South America.** The striped green area showcases the physical boundaries of the Amazon rainforest. The area in orange shows regional dependence on precipitation originating within the Amazon rainforest, with darker shades indicating higher dependency.

The electricity generation potential of hydropower plants is thus likely to be negatively impacted by longer periods of drought due to a reduction in precipitation patterns (Shu et al., 2018). Investors and national governments can minimize the risk of stranded assets by mapping which hydropower plants are at risk of such abrupt shifts in precipitation.

Our initial analysis shows that 998 hydroelectric powerplants (documented as being either in construction or with planning approved in South America), constituting a total of 66.97 GW, are most likely to be affected by decreases in precipitation in four countries (Brazil, Argentina, Peru and Bolivia) and within eight river basins (see Figure 2). Out of the 998 dams identified in risk regions, 65 of the largest dams, accounting for about 64% (42.61gW) of future power generation, are located in Brazil (26), Peru (24), Bolivia (11) and Paraguay. Although the extent to which electricity production will be affected by extreme weather events will differ, a disruption to the hydrological cycle increases the risk of power plants to become stranded assets.

Out of the 65 hydropower projects under analysis, 15 were under the control of a consortium with at least one publicly-listed company, bringing the total to 10 different publicly-listed companies investing in these projects. Nevertheless, the financial implications extend beyond the region. For example, 63% of the 548 are located in the Global North, with US, Spain, Great Britain, Canada, France, Germany, Switzerland and Bermudes being represented most. A total of 438 total unique shareholders, with at least 0.01% of ownership, could have investments at risk, including banks (131), individuals or family groups (127), corporation companies (103), mutual pension funds and trustees (58), finance companies (53), public authorities, state institutions and governments (29), insurance companies (29) and foundations or research institutes (6).

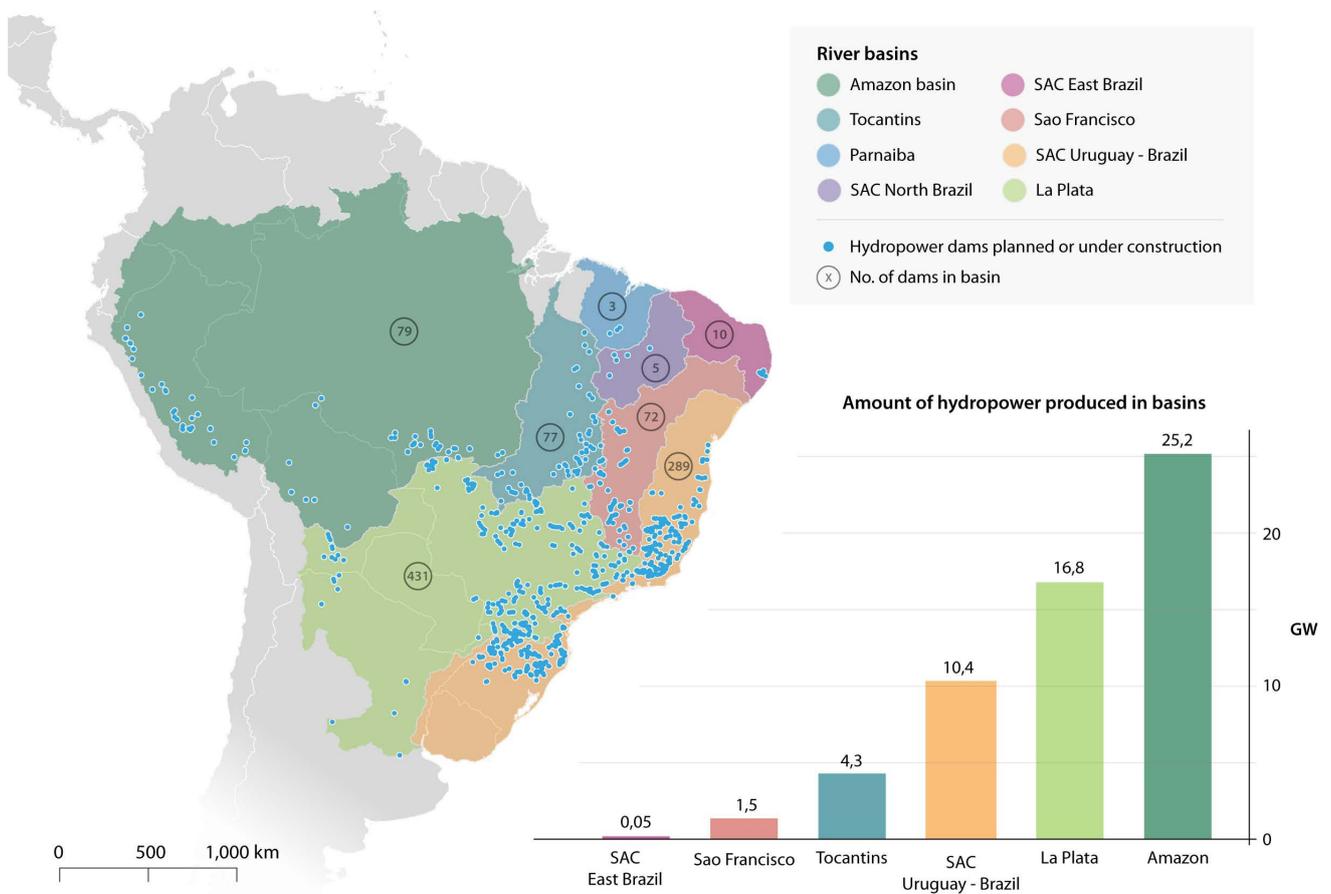

**Figure 2 | Hydropower stations in constructions or in planning across basins at risk of decreasing levels of precipitation.** The map showcases the main river basins which are likely to be impacted by changes in the hydrological cycle as a result of a tipping of the Amazon rainforest. The dots mark the location of all hydropower dams planned or under constructions within the boundaries of these basins.

The top 30 unique shareholders own about 80% of the $169.14bn in total equity, of which eight are state agencies and the governments, totaling $43.63bn in shares. Large US-based asset managers (The Capital Group Companies Inc., BlackRock Inc., Vanguard Group, and State Street Corporation) own 18% of the total equity size and have investments across several of the 10 publicly listed companies under analysis. Notably, the Brazilian government controls more than 48% of a few small companies, while another four unique shareholders (SAS Rue La Boetie, State Street Corporation, JPMorgan Chase & Co, The Government of Canada) are found owning shares in at least six of the publicly listed companies, although the investments is comparatively small, less than $3bn.

Here we have shown the risk of investments becoming stranded assets due to a possible transgression of a tipping point in the Amazon, and its impacts on precipitation patterns at the regional scale. Nevertheless, this analysis has broader implications for investors. Physical risks are highly relevant for the owners of hydropower plants in the affected regions, while potential changes in public awareness and regulation can represent high transition risks for investors who fund deforestation-linked activities. Furthermore, many of the large hydropower plant projects have been opposed by indigenous and local communities due to the high social and ecological costs such projects would have incurred (Mapstone, 2011). Furthermore, as the recent draught in Manaus has shown, prolonged draughts, which can be the effect in hydrological cycle, have impacts beyond electricity generation, extending to agriculture and communities dependent on riverine routes for supplies (Knutson, 2023). The social and financial risk incurred by lower precipitation levels further emphasize the need for policies and enforcement in Amazon countries that limit deforestation activities. Financial institutions have, as we have shown, a clear direct incentive to support this work.

## Chapter 7.

# The potential of 'nature tech' for Latin America and the Caribbean

*Svante P. Persson, Regina Cervera and Constanza Gomez Mont*

Can you imagine a world where technology and nature thrive together? In a world grappling with climate change and biodiversity loss, transformative advancements utilizing a range of new technologies are drawing the interest of those at the forefront of efforts to protect Earth's diverse ecosystems.

The potential of nature tech (Nature4Climate, 2022) – defined as technologies aimed at addressing significant environmental issues like climate change and biodiversity loss, as well as those that facilitate the enhancement of nature-based solutions (NbS) – stands as a potential powerful game-changer for preserving our planet's rich biodiversity and tackling the challenges of climate change head-on. How? By filling essential data gaps (Sistemiq, 2022) for transparent, effective, and trustworthy science-backed strategies and policies. These innovations empower us with comprehensive information crucial for making wise decisions in land use planning, management, conservation and restoration. As the global economy increasingly acknowledges the need for sustainable practices, businesses and projects that embrace nature tech are more likely to thrive.

Additionally, these cutting-edge tools have the potential to supercharge the expansion of smarter, eco-friendly, low-carbon agriculture (Van Acker, Fraser and Newman, 2022) and nature-positive supply chains (Bloomberg, 2023). Not only does this make our efforts more environmentally responsible, but the added transparency and resilience thanks to a solid foundation of data, will also draw in a wave of new investors eager to support sustainable initiatives (Milborrow, King and Bromfield, 2023), as it has the potential to reduce uncertainty and risk associated with environmental projects and assures investors that their funds are being used efficiently and for the intended purpose. Moreover, nature tech enhances our ability to make informed decisions about land use planning, management, conservation, and restoration. When investors see that technology is optimizing the allocation and utilization of resources, they become more inclined to invest in projects that promise sustainable outcomes.

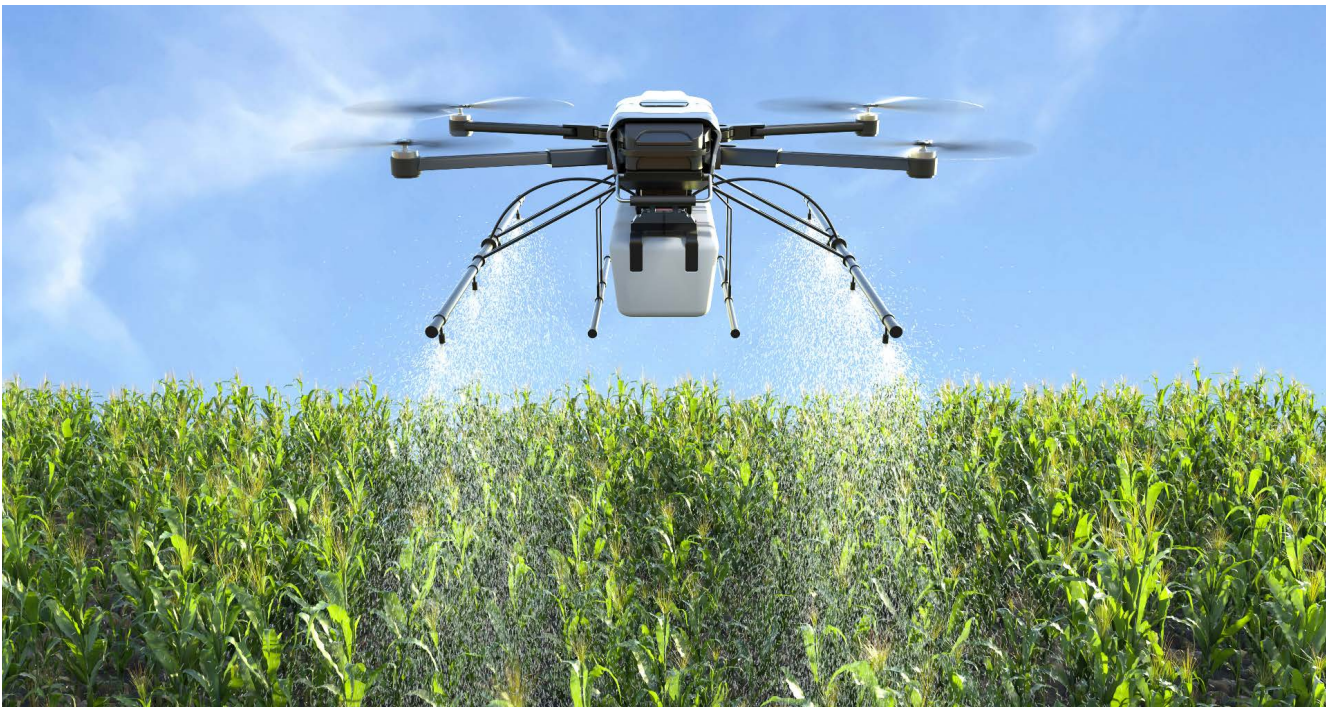

## The nature tech landscape in the Latin America and the Caribbean (LAC) region

This is particularly relevant in biodiversity-rich regions like Latin America and the Caribbean (Watson, Debade and Gallego, 2023), which collectively holds 40% of the world's biodiversity, 30% of its freshwater resources, and nearly 50% of its tropical forests. As these technologies, such as machine learning, eDNA, genomics, and networked sensors, advance, they hold the promise of accelerating data-based decision-making and effective local action in the rapidly evolving field of biodiversity conservation and regeneration.

However, as we navigate this transformative landscape, it becomes evident that challenges loom large (Speaker et al., 2021) to advance these innovations in the region: limited funding and allocation where it is needed the most, a need for a wide-spread access to context-based tools and technologies, a disconnected ecosystem of collaboration, and inadequate capacity building represent hurdles that must be surmounted for the region to fully take advantage of the nature tech momentum. Moreover, as we venture into the transformative landscape of nature tech, it's essential to adopt ethical approaches that promote responsible technology usage. These approaches not only ensure that the unintended consequences and over-expectations of new technologies are mitigated but also that the identified vulnerabilities are addressed. Instruments, such as the UNESCO's Recommendation on the Ethics of Artificial Intelligence (UNESCO, n.d.) offer pathways to rethink the development of AI systems and carry out more sustainable and ethical projects. For example, a crucial aspect of ethical technology deployment is the rigorous evaluation of its direct and indirect environmental effects. This encompasses assessing the carbon footprint of AI systems, which, during their training stage, can emit anywhere from 25 to 500 metric tons of $CO_2$ (Hao, 2019). Moreover, embracing ethical approaches means ensuring that technology benefits all segments of society. This means actively involving communities and indigenous peoples, as well as women and youth, throughout the entire life cycle of these systems. By fostering inclusivity, we not only honor the rich diversity of perspectives of the LAC region, but also make certain that technology solutions are representative of various cultures, contexts, and the unique needs of different communities.

Fortunately, the path forward also offers opportunities to unlock this potential: increasing radical and empathetic multi-sector and transboundary collaboration, improving the interoperability of data and technologies to facilitate information sharing, the application of frameworks of responsible uses of new technologies, as well as enhancing tech literacy and capacities for large-scale environmental data analysis.

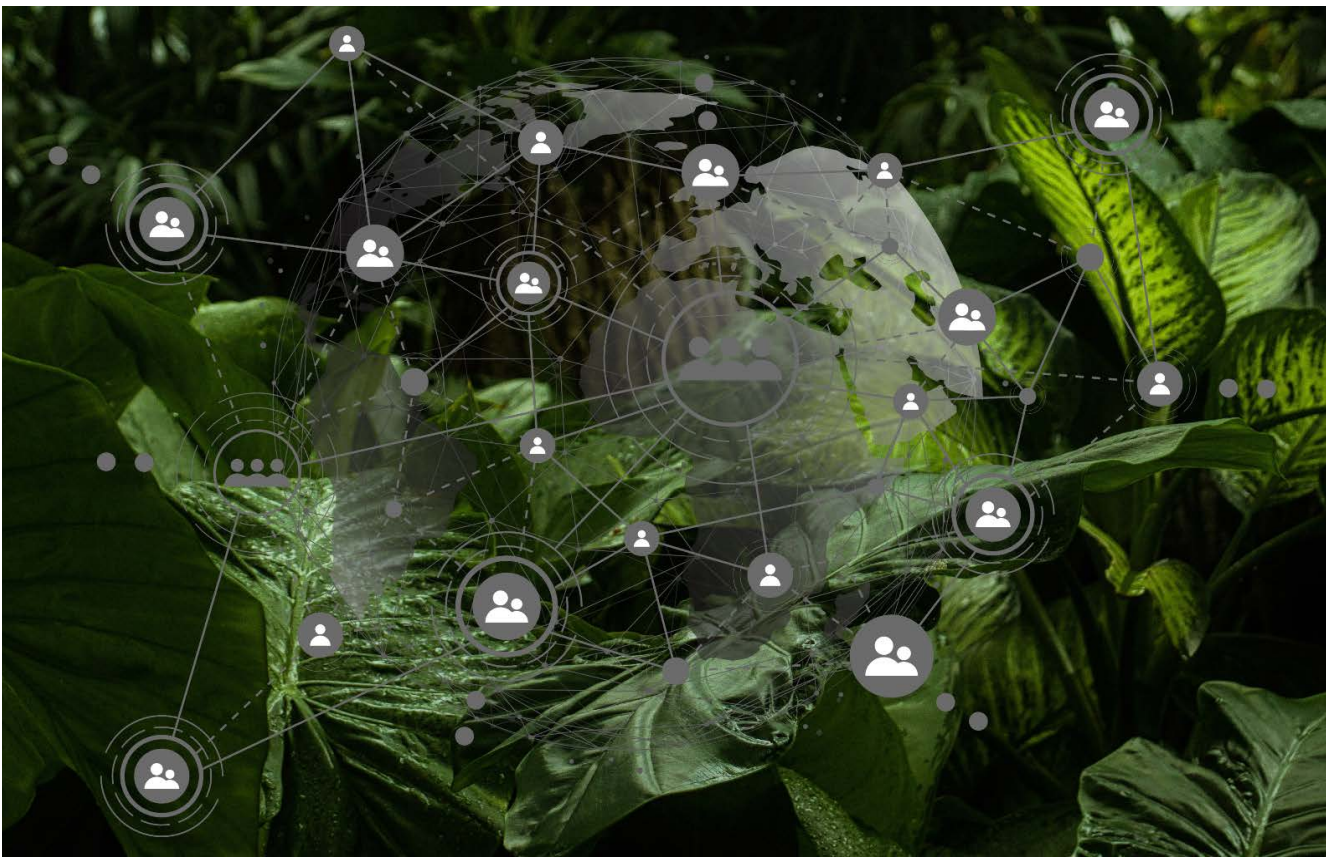

As we address these challenges and grasp these opportunities, we can catalyze a shift in the investment landscape, unlocking the potential of nature tech.

## A shift in the investment landscape and the rise of new opportunities

There is now a rising acknowledgment of nature's significance as a fundamental infrastructure asset for businesses (WEF, 2020), and we can expect an increased desire for nature-focused technologies as the corporate and finance sectors pivot towards nature-positive strategies. As we navigate this evolving landscape, it is essential to recognize that capital flows to nature at the speed of trust, and nature tech represents a new market (Eng, King and Strong, 2022), critical for building trust in nature-based solutions.

As disclosure regulations are recommended by the Taskforce on Nature-related Financial Disclosure (2023), nature tech could help meet the growing demand for transparency and monitoring throughout value chains. Notably, there are innovative solutions emerging from the Global South (Economist Impact and JP Morgan, 2022) that are actively tailored to the needs of smallholders and local communities, and some of the most significant innovations in nature tech are expected to originate from the Global South (Lema and Rabellotti, 2023).

## What's next?

Nature tech is not just a trendy catchphrase; it embodies a profound paradigm shift in our approach to biodiversity conservation and restoration. As the world grapples with the urgent call for government action in combating climate change and biodiversity loss, we must not overlook the equally critical role of private sector investment in expanding nature-based solutions.

Yet, the clock is ticking. We have precious little time to align ourselves with global targets and secure a future that's not just habitable but flourishing and sustainable. In the face of these pressing challenges, the imperative for international collaboration and data sharing becomes all the more apparent.

To fully unlock the immense potential of nature tech, we must act fast and decisively. This means rapidly building capacities, fostering radical collaboration that transcends borders, and creating platforms for the exchange of knowledge, ideas, and expertise. Most importantly, we need to catalyze the responsible adoption of new technologies, ensuring that they are tailored to the needs of the most vulnerable populations and follow a human rights-based as well as an ethical approach.

### Innovations are on the rise in the LAC region

The *NaturaTech LAC* initiative led by the IDB Lab and local partners aims to enhance the effectiveness, scalability and transparency of biodiversity conservation and regeneration action, as well as sustainable finance to catalyze investments in this field, by fostering the development and adoption of new technology solutions in the LAC region, while fostering North-South and South-South knowledge and resource collaborations.

The mission of this initiative involves accelerating the adoption of new technologies, testing novel models, and providing funding, including innovative financing approaches, with a special focus on projects that benefit local communities and that integrate indigenous peoples' perspectives. By doing so, it seeks to usher in innovative approaches that expedite the achievement of global commitments and targets, such as the Global Biodiversity Framework (CBD, 2022). Importantly, this initiative recognizes the importance of understanding local needs, contexts, and the diverse nature of the LAC region, translating these insights into tangible action.

Stay tuned for more news on this initiative at bidlab. org.

## Chapter 8.

# Co-creating natural capital solutions in the Amazon Basin with communities, governments, and financial leaders

*Adrian L. Vogl and Mary Ruckelshaus*

The economic and fiscal consequences of inaction on loss of biodiversity and the ecosystem services it provides are quite significant (OECD, 2019), given that US$44 trillion of global GDP—around half—is highly or moderately dependent on nature (WEF, 2020). Research shows that protected areas are not sufficient to maintain biodiversity and its benefits—especially in the face of climate change -- thus, finding common visions for management and investments in production landscapes and seascapes is crucial to securing sustainable livelihoods and human wellbeing (Huntley, 2014).

Despite growing recognition that biodiversity protection is fundamental to achieving food security, poverty reduction, and sustainable development, its value is not widely integrated into decision-making processes for public policies and investments. However, Natural Capital Assessments and Accounting (NCAA) are proving to be effective approaches for integrating information about the benefits of biodiversity and ecosystem services, drawn from diverse sources of knowledge and understanding, into policy and finance decisions (Natural Capital Project, 2023). Natural capital approaches are designed to directly inform decision-making, integrating people and the planet's life-support systems into economic development (Ruckelshaus et al., 2022). Natural capital approaches are most effective when embedded in a science-policy process driven by both technical and policy experts, who are knowledgeable about a country or region's priorities, as well as which policy and finance interventions are possible and relevant within the specific socio-political context. Strong stakeholder engagement is key to ensuring diverse sources of knowledge and values are considered.

Natural capital approaches are already impacting policy and investment in the Amazon Basin. In this chapter, we highlight a handful of examples of such impact, illustrating the value of participatory science-policy processes where NCAA approaches inform policy and investment decisions for the benefit of multiple sectors and communities. A growing community of practice—informed by technical, policy, and finance experts from the region and around the world—is poised to further develop capacity and demonstrable benefits for Amazonian people, their prosperity, and the planet.

## Community resilience in the Amazon headwaters of Peru, Bolivia, and Brazil: Reducing risk from drought, floods, water pollution, and mosquito-borne disease

The tri-national area in the states of Madre de Dios (Peru), Acre (Brazil) and Pando (Bolivia) in the southwestern Amazon is a biologically, culturally, and socially diverse landscape that is increasingly vulnerable to extreme drought, flooding, and zoonotic diseases (see Chapter 3). These threats are driven by a combination of changing climate and human-driven disturbance in the form of rapid and unplanned urban development and extensive deforestation for gold mining, cattle ranching, and subsistence farming. To help address these pressing issues, the Natural Capital Project at Stanford (NatCap) worked with the Regional Government of Madre de Dios in Peru, municipal leaders and environmental planners in the region, the Amazon Center for Scientific Innovation (CINCIA), Cayetano Heredia University of Peru (UPCH), and Herencia in Bolivia to co-create actionable, science-based information on the values of natural capital to inform key policy opportunities.

Through a process of co-development, the project team identified critical challenges around water security and helped describe possible futures for the region consistent with stakeholders' expressed visions and values. The results demonstrated strong linkages between land clearing and land use change, and downstream impacts on water resources, flood risk, and incidence of mosquito-borne diseases such as malaria and dengue (Guevara et al., 2020). New analyses of vulnerability to flooding revealed how flood risk and communities' adaptive capacities could be affected by future land use changes. These findings (PRO-Agua, n.d.) helped make the case to policymakers at regional and municipal levels for forest protection and management through the MERESE-Hídrico (payments for watershed ecosystem services) framework (Guevara et al., 2020).

As a result of this work, regional government decision-makers and planners incorporated freshwater ecosystem services as key selection criteria in the design and implementation of watershed and land use management plans, two new MERESE-Hídrico areas were identified, and a committee was institutionalized within the Madre de Dios regional planning framework to streamline the identification and adoption of such programs in the future. Furthermore, some of the legal barriers to these payments for watershed services programs were addressed by codifying MERESE-Hídrico projects as an eligible category of protected areas within the Regional System of Protected Natural Areas.

## Harmonizing livelihoods, cultural heritage, and water resources in the Llanos de Moxos region in Beni, Bolivia

The complex landscape of the Llanos de Moxos (LdM) is the result of its unique position as the largest seasonally flooded savanna ecosystem in the Amazon, its history of more than 10,000 years of human occupation, and its wealth of biological and cultural diversity. This extensive system of wetlands-savannas-forests is home to approximately half a million people including numerous Indigenous peoples, whose livelihoods largely depend on direct use of natural resources and the integrity and functionality of the ecosystems that sustain them. Rapid expansion of mechanized agriculture, alluvial gold mining, fires, and overexploitation of timber, fisheries, and wild game is putting pressure on this unique landscape and its ability to provide critical ecosystem services to local communities as well as those downstream in the Amazon Basin (Vogl, 2022).

NatCap collaborated with a coalition of local and international researchers and NGOs[*] to engage stakeholders at the Departmental and Municipal levels, generate new science, and build capacity for policymakers to integrate the benefits derived from the ecosystem services of this region into policy design and development planning.

---

\* The Working Group for the Llanos de Moxos (CIBIOMA, n.d.) includes Armonía, the Center for Research in Biodiversity and Environment at the Autonomous University of Beni (CIBIOMA), FaunAgua, University of Bonn, Gordon and Betty Moore Foundation, Stanford University's Natural Capital Project (NatCap), and the Wildlife Conservation Society (WCS).

The project partners collaborated to develop a comprehensive assessment of livelihood systems in the Llanos de Moxos. They articulated alternative development scenarios that highlighted conflicting visions for the future of the region and illuminated their impacts on key ecosystem services and human wellbeing (Vogl et al., 2022). Comprehensive mapping of actors influencing land use and development policy in Beni was used to design engagement and outreach. As a result, local leaders in each of the 19 municipalities in Beni Department have specific, locally supported, and actionable social-ecological and spatial information on the impacts of climate change and land transformation on ecosystem services and communities, enabling them to make decisions and to design development plans incorporating the potential of this landscape to boost the well-being of their constituencies.

## Natural capital assessment helps target priority investments in Colombia

Colombia is home to an estimated 10 percent of all species in the world and has the highest diversity of birds and orchids globally. The Inter-American Development Bank (IDB) and the Colombian government invited NatCap to collaboratively develop a nationwide natural capital assessment to identify biodiversity and ecosystem-service hotspots. These hotspots are now part of the IDB's Country Strategy for Colombia, used to guide priority investments by the government and IDB for protection, stewardship, and revitalization. The assessment was designed to answer two questions: what areas of the country provide the greatest benefit to people in Colombia, across multiple ecosystem services; and to what degree do these ecosystem service hotspots fall within existing protected areas and indigenous reserves?

The team used the InVEST suite of models (Natural Capital Project, 2023) to spatially quantify four key ecosystem services: (i) climate regulation, in terms of carbon stored by ecosystems; clean water, in terms of both (ii) sediment retention and (iii) nitrogen retention for people downstream; and (iv) coastal protection, in terms of reducing risk to people in coastal areas from flooding and erosion.

The results showed that the Amazon region of Colombia contains the greatest biodiversity of any other region nationally. Forests in the Amazon (and Pacific) regions are especially important for carbon storage. The shared investment strategy for IDB and Colombia is now based on a clear picture of how targeted investments in the Amazon region will generate the greatest returns for biodiversity and climate mitigation.

## The potential of natural capital approaches in the Amazon

In addition to working with communities in the headwaters of the Amazon to understand their challenges and priorities, natural capital approaches were used to show the connections between upstream forest management and downstream water security, flood risk, and zoonotic disease. This facilitated the establishment of payment for ecosystem services programs that incentivize protecting and restoring nature to preserve the services it provides. In the Llanos de Moxos in Bolivia, the largest seasonally flooded savanna ecosystem in the Amazon, natural capital approaches provided information for local leaders to design development plans that maximize benefits to both livelihoods and ecosystems. In Colombia, a nationwide natural capital assessment is now guiding investment in key areas to maximize benefits.

Natural capital approaches are now being used in a variety of contexts to help connect the dots for decision-makers between how people value and manage their ecosystems, and the returns they see from those ecosystems – like water security, flood protection, disease mitigation, and livelihoods. Policymakers across the Amazon Basin are increasingly recognizing that strategic development and targeted investments in such benefits make their communities stronger in a multitude of ways, especially when participatory science-policy processes are used. These projects bring stakeholders together to co-develop a vision as well as the science needed to inform specific mechanisms and policies, moving them closer toward a future where people and nature can thrive.

*Chapter 9.*

# Water connections: sustainable futures for the Brazilian Amazon, Cerrado, Caatinga, and semiarid regions

*Ana Paula Dutra Aguiar, Taís Sonetti-González, Minella Martins, Francisco Gilney Bezerra and Aldrin Perez-Marin*

The Brazilian semiarid region has more than 31 million inhabitants and occupies an area of over 1 million km², including the entire Caatinga biome and portions of the Cerrado biome (Figure 1.a). Running through biomes, the São Francisco River basin is of crucial economic, ecological, and cultural importance to the region. The river is also fundamental for the supply of water, food, and energy to the region and to the world through the production of irrigated agricultural commodities (Bezerra et al. 2019). The region, in particular the Caatinga biome, also has enormous potential for the expansion of renewable energies, in particular wind and solar plants (Neri et al. 2019). The expansion of large-scale projects for food and energy production has reshaped the region, presenting both opportunities and risks.

Particularly in the Cerrado part of the region, aggressive deforestation, loss of biodiversity and depletion of water resources due to irrigation projects are some of the environmental consequences of this process (Strassburg et al. 2017; Silva et al. 2021; De Espindola et al. 2021). For example, official information shows that 72% of the water is withdrawn for irrigation (ANA, 2021), leading to conflicts related to water use (CPT, 2021, Peixoto et al. 2022; Martins et al., in prep) (Figure 1.b). On the other hand, the overexploitation of water resources also brings distal impacts. For example, the Cerrado and the Amazon, though seemingly distinct, constitute a dynamic unit due to the water cycle interactions (Costa and Pires 2010). These connections establish the Cerrado as a vital balancing system for the regional climate. Furthermore, the Cerrado plays a pivotal role because it contains the headwaters and the largest portion of South American watersheds (Latrubesse et al. 2019).

The semiarid is also an area of great socioeconomic contrasts - which largely reproduces the multiple inequalities that still characterize Brazilian society (Figure 1.c). All municipalities in the semiarid region have a Human Development Index (HDI) lower than that of Brazil, an index that takes into account indicators of longevity, education and income (IPEA, 2023). However, although historically stigmatized as an impoverished region ravaged by drought (Alvalá et al. 2019), the semiarid population through a strong network of social movements have outlined new perspectives for the future, notably through the paradigm of "coexistence with the semiarid" (Pérez-Marin et al. 2017). The new paradigm is based on access to water and land, as well as the adoption of practices of traditional agroecological production. This new paradigm coexists - not without conflicts - with the expansion of mega-enterprises for food and renewable energy production, as the growing prominence of commodities in the country's economy allied to the deindustrialization process in the last decades (Nassif, Bresser-Pereira, and Feijo 2018; Fonseca, Arend, and Guerrero 2020) reinforced the political power of the primary production sectors (Ioris, 2016, Rochedo et al. 2018).

In this complex situation, the challenges to achieving a sustainable and fair future are enormous. Historical conflicts and inequalities have led to huge power imbalances and hindered any meaningful dialogues regarding a sustainable future, in which multiple voices and perspectives could be taken into consideration. To address this situation, the XPaths project (www.xpaths. org) undertook an extensive two-year participatory process, with the participation of more than a hundred stakeholders from several sectors and acting at different spatial scales. The process combined state-of-the-art multiscale participatory and system thinking approaches (Collste et al. 2023; Nguyen and Bosch 2013; Aguiar et al., 2023; Aguiar et al., in prep). The culmination of this effort was the identification of four primary challenges and the formulation of four interconnected strategic actions aimed at their resolution (XPaths, 2023), which we summarize below.

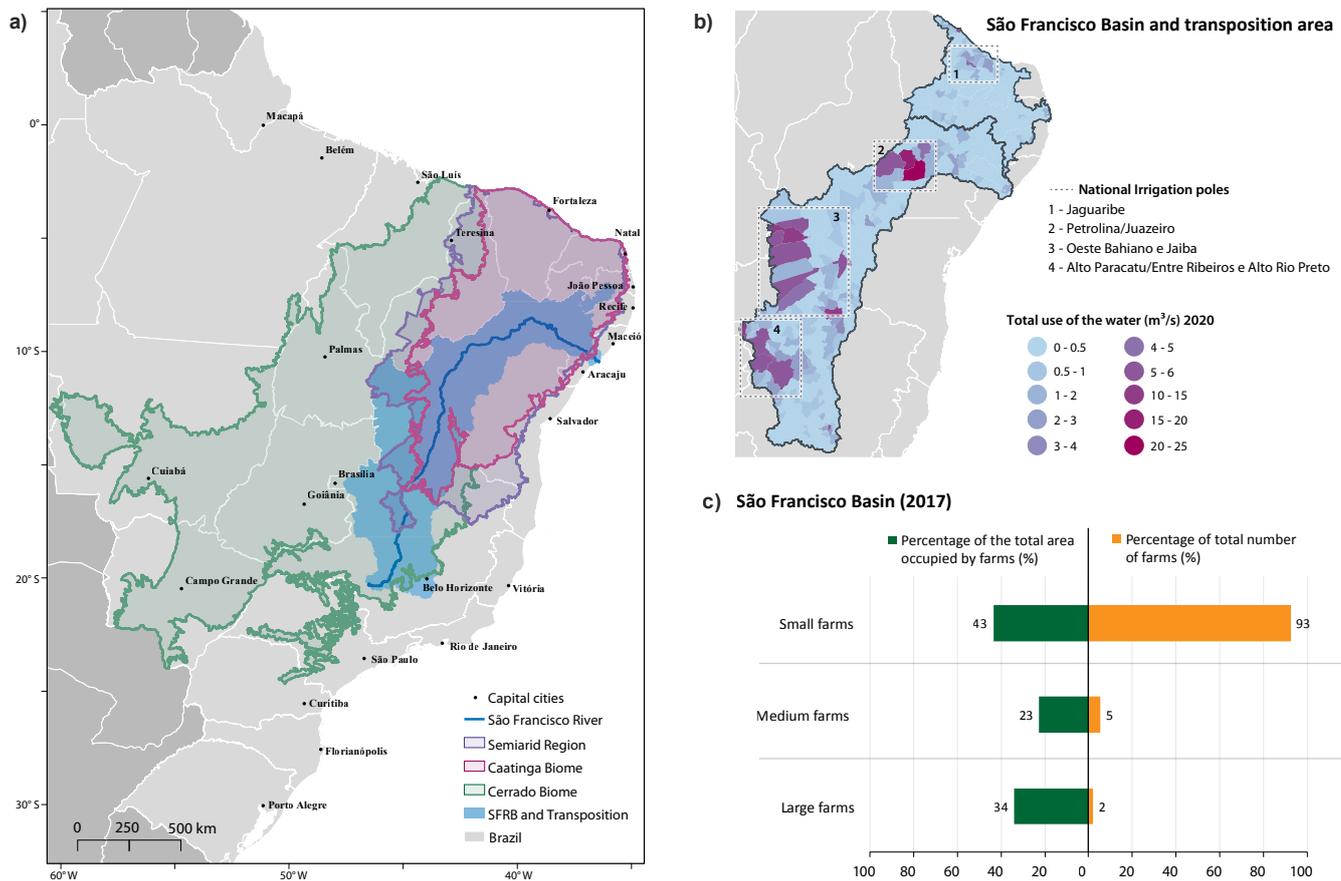

**Figure 1** | (a) Location of the semiarid and São Francisco River Basin in Brazil and their location in relation to the Cerrado and Caatinga biomes (prepared by the authors); (b) Map illustrating how the municipalities in irrigation poles (the four black rectangles) consume most of the water in the region (Source: adapted from Martins et al. (in prep) using ANA (2021)); (c) Graph illustrating the land concentration in the Sao Francisco River. Note how the large farmers represent only 2% of the number of farms, but occupy 34% of the land, while the small farmers are 93% of the number of farms, but occupy only 43% of the land (Source: adapted from Martins et al. (in prep) using IBGE (2017)).

## Strategic Actions: Challenges and Solutions

- **Strategic Action 1:** The first challenge prioritized by the stakeholders is the impact of large agricultural, energy and mining projects on the quality and quantity of water in the rivers of the São Francisco River Basin, including unequal access to the resources and resulting divided conflicts. The proposed solution was the implementation of a broad Environmental Education, Communication, and Social Mobilization Program for the São Francisco River Basin, to create consciousness about the multiple uses of the water. Participants named the program: "ÁGUA VIVA - Building a biodiverse environmental awareness".

- **Strategic Action 2:** The second challenge is the historical land concentration which causes unequal access to land and, consequently, to water. The proposed solution is the implementation of an agrarian reform compatible with the traditional practices of the biomes and the demarcation of territories of traditional peoples and communities. Agrarian reform is the necessary basis for Strategic Action 4.

- **Strategic Action 3:** The third critical problem identified by the participants is the concentration of political power in the hands of the dominant economic sectors such as agribusiness and mining sectors, a historical problem reinforced by the current cycle of expansion of commodities in the region. Derived from this is the discontinuity of public policies aiming at the public good, hindering the region's sustainable development. The proposed solution is a broad political training program to increase social awareness and social responsibility, promoting participation, leadership and political change.

- **Strategic Action 4:** Finally, the Brazilian economy's dependency on commodities and deindustrialization were pointed out as a critical structural problem influencing the region. Strategic Action 4 is therefore broader in scope, but complementary to others. The participants proposed a series of actions at the national and international levels to foster a new development model. The aim would be to help diversify the economy and improve social well-being. Examples of proposed actions at the national level are investments in agroecological food production aiming at increasing food sovereignty, support to small-scale farmers, investments in reindustrialization, capacitation and training, policies for inequalities reduction and strengthening of legal frameworks; at the international level: monitoring of the socio-environmental impacts of commodities production (beyond deforestation), including water availability/pollution, conflicts, land as a financial asset and dispossession; communicate the multiple socio-environmental impacts of the commodity chains for investors and international markets; review international frameworks and agreements (e.g., ILO Convention 169, Mercosur-EU Agreements, etc.).

These strategic actions encompass multiple Sustainable Development Goals (SDGs), aligning with the integrative and universal spirit of the 2030 Agenda. Our approach identifies actions that attempt to address the systemic structures and deep roots of unsustainability (Aguiar et al., in prep). These actions, however, can only be implemented through the collective commitment and engagement of diverse stakeholders across various levels. Understanding is needed across national and international platforms, engaging a wide array of actors. The successful execution of these actions stands as a vital safeguard for the semiarid region, preserving the water source not only for Brazil but for a significant portion of South America.

# Chapter 10.
# Conclusion

*Megan Meacham and Victor Galaz*

The Amazon region has been long recognized for its ecological diversity, importance for the climate system, and deep significance for people and indigenous communities in the region. Over time, we have come to understand this iconic biomes' importance for local communities, Latin American countries' economies, and the possibilities for a prosperous future for all. In a globalized world, what happens in the Amazon does not stay in the Amazon. Protecting its resilience for the future in ways that are ecologically viable but also socially just, is an issue of global importance and concern.

This report has shown the various ways that humans, ecosystems and the climate system are interwoven, based on the latest insights from the resilience and sustainability sciences. While this close interconnection between the social and ecological might sound obvious, a growing body of research that explores the connections between economic activities, financial investments, and social-ecological change consistently shows system fragilities and risks that remain unaddressed by key decision-makers in the public and private sector.

Economic development depend on the functioning of the biosphere, but are eroding long-term sustainability by threatening biodiversity and altering ecological processes, all with unequal impacts on people in the region. Businesses and financial actors depend on stability and predictability, however, there are large risks of abrupt and irreversible changes that would fundamentally alter the material foundations for economic activities, and limit the predictability for future investments. The Amazon region is changing at unprecedented speed, forcing us all to rethink the role of public and private investments.

In 2022, we presented a simple framework to assess the role of finance to support a just transition towards sustainability. We believe that this framework is just as applicable when discussing the future of the Amazon.

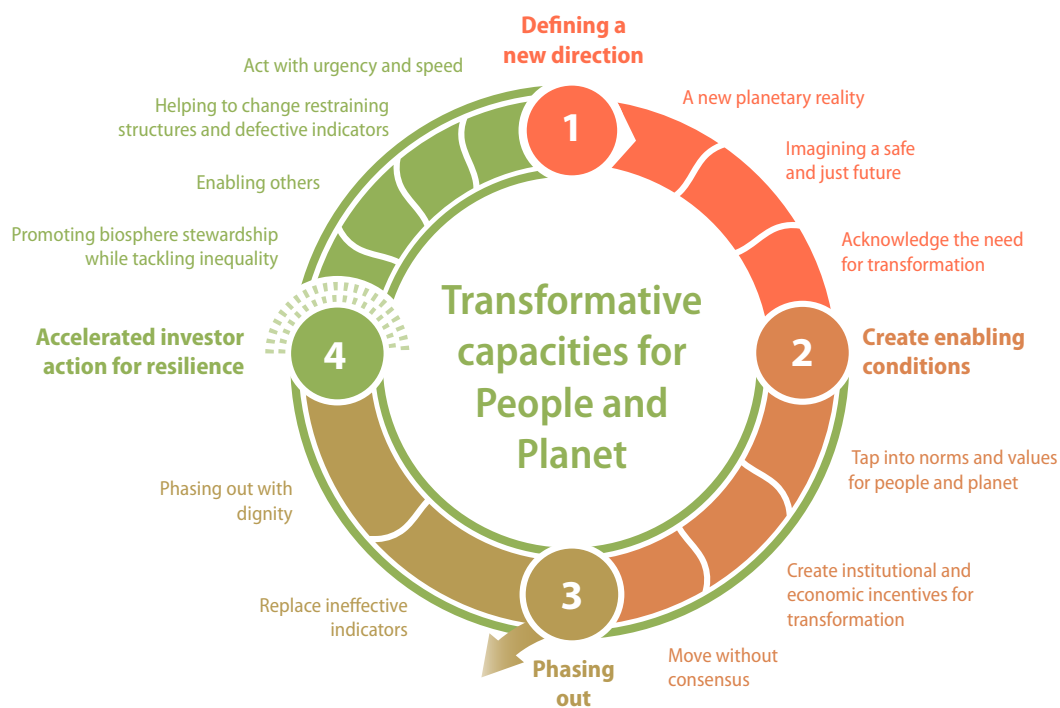

A plan for finance to support resilience and a just transition for the Amazon needs to acknowledge its rapidly changing reality, including possible abrupt and cascading changes (Chapters 4 and 5). It also needs to consider the multiple cross-national connections between investors (Chapter 3), and challenging systemic risks for investors that evolve as a result of loss of the web of life (Chapter 2), and shifts in major biosphere dynamics such as precipitation patterns (Chapter 6).

Fortunately, new technologies (Chapter 7) and new science-based approaches such as natural capital assessments (Chapter 8) are helping to understand these processes and the investments, policies and regulations that will ensure their sustainability to both people and planet. There is an urgent need to expedite the use of such tools that both help phase out harmful investments, and support accelerated investor action for resilience. This alignment of financial actor interests and the needs for environmental protection is an opportunity for both governments and private investors to invest in 'nature positive' projects. The ambition can no longer be to solely limit damages, but restoration and regeneration will be necessary to stabilize the resilience of the Amazon biome.

Recognizing the interconnectedness of ecosystems and human communities, it is imperative to prioritize the involvement of local communities, especially indigenous groups, in decision-making processes. Respecting and protecting the rights of these communities is not only ethically sound but also vital for the overall success of initiatives like "Amazonia Forever". As our colleagues have shown (Chapter 8, and 9), that social equity goes hand in hand with strategic investment decisions.

Sustainable investments aligned with responsible practices can pave the way for a just and resilient Amazon, contributing to the well-being of current and future generations.

There is still time, yet no time to waste.

Beijer
Institute
OF ECOLOGICAL ECONOMICS

KUNGL.
VETENSKAPS
AKADEMIEN
THE ROYAL SWEDISH ACADEMY OF SCIENCES

Stockholm
Resilience Centre

Stockholm
University

MISTRA
FinBio
FINANCE TO REVIVE BIODIVERSITY

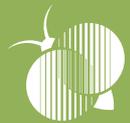

natural
capital
P R O J E C T